\documentclass[aps,twocolumn,superscriptaddress,amsmath]{revtex4-1}

\pdfoutput=1

\usepackage{dcolumn}
\usepackage{bm}
\usepackage{amsfonts}
\usepackage[pdftex]{color,graphicx}
\usepackage{amssymb}
\usepackage{rotate}
\usepackage{subfigure}
\usepackage{subfigure}
\usepackage{amsmath}
\usepackage{amsthm}
\usepackage{hyperref}

\newcommand{\ket}[1]{\ensuremath{\left| #1 \right\rangle}}
\newcommand{\bra}[1]{\ensuremath{\left\langle #1 \right|}}
\newcommand{\sand}[2]{\left\langle #1| #2\right\rangle}

\newcommand{\f}[2]{{\ensuremath{\mathchoice%
       {\dfrac{#1}{#2}}
       {\dfrac{#1}{#2}}
       {\frac{#1}{#2}}
       {\frac{#1}{#2}}
       }}}

\renewcommand{\-}{\,-\,}

\newcommand{\avg}[1]{\left< #1 \right>}

\newcommand{\set}[1]{\{ #1 \}}
					
\let\oldmarginpar\marginpar
\renewcommand\marginpar[1]{\-\oldmarginpar[\raggedleft\tiny #1]%
{\raggedright\tiny #1}}

\newcommand{\argmin}{\operatornamewithlimits{arg\ min}}

\graphicspath{{figs/}}

\begin{document}


\title{Approximating random quantum optimization problems}
\hypersetup{pdftitle={Approximating random quantum optimization problems},
	pdfauthor={B. Hsu, C.R. Laumann, R. Moessner, and S.L. Sondhi}}
	

\author{B. Hsu}
\affiliation{Department of Physics, Princeton University, Princeton, NJ 08544}
\affiliation{Max Planck Institut f\"ur Physik Komplexer Systeme, %
	01187 Dresden, Germany}

\author{C. R. Laumann}
\affiliation{Department of Physics, Harvard University, Cambridge, MA 02138, USA}

\author{A. M. L\"{a}uchli}
\affiliation{Institute for Theoretical Physics, University of Innsbruck, 6020 Innsbruck, Austria}

\author{R. Moessner}
\affiliation{Max Planck Institut f\"ur Physik Komplexer Systeme, %
	01187 Dresden, Germany}

\author{S. L. Sondhi}
\affiliation{Department of Physics, Princeton University, Princeton, NJ 08544}

\date{\today}


\begin{abstract}

We report a cluster of results regarding the difficulty of finding approximate ground states to typical instances of the quantum satisfiability problem $k$-QSAT on large random graphs.
As an approximation strategy, we optimize the solution space over `classical' product states, which in turn introduces a novel autonomous classical optimization problem, PSAT, over a space of continuous degrees of freedom rather than discrete bits.
Our central results are:
(i) The derivation of a set of bounds and approximations in various limits of the problem, several of which we believe may be amenable to a rigorous treatment.
(ii) A demonstration that an approximation based on a greedy algorithm borrowed from the study of frustrated magnetism performs well over a wide range in parameter space, and its performance reflects structure of the solution space of random $k$-QSAT. Simulated annealing exhibits metastability in similar `hard' regions of parameter space.
(iii) A generalization of belief propagation algorithms introduced for classical problems to the case of continuous spins. This yields both approximate solutions, as well as insights into the free energy `landscape' of the approximation problem, including a so-called dynamical transition near the satisfiability threshold. 
Taken together, these results allow us to elucidate the phase diagram of random $k$-QSAT in a two-dimensional energy-density--clause-density space.

\end{abstract}

\maketitle

\tableofcontents


\section{Introduction} 
\label{sec:introduction}

Ever since the realization that quantum mechanics could be used to solve computationally difficult problems, there has been an intense effort to understand quantum computers and quantum algorithms.
Naturally, much of the algorithmic research front has been focused on those that give a substantial speed up over their classical counterparts.
Complexity theory suggests, however, that there are many natural optimization problems that are hard even for a quantum computer to solve.
These include both classical NP-complete problems such as satisfiability (SAT) and intrinsically quantum QMA-complete problems such as quantum satisfiability
(QSAT) which require a quantum computer even to represent the relevant state space.
These hardness results suggest the need for approximate strategies where one settles for less-than-globally-optimal solutions.
Unfortunately, this too may be hard as evidenced by the classical PCP theorem \cite{Arora:1998,Arora:2009zv}.

The PCP theorem, in this setting, implies that there are classical energy functions for which it is NP-complete not only to find the ground state but even to find highly excited states with an energy \emph{density} below some threshold.
This result has astonishing dynamical consequences. Under the assumption that nature cannot solve NP-complete problems, there are \emph{no} dynamics which can relax these systems to low temperature.
This constitutes an abstract argument for the existence of glassiness which, in an amusing parallel with the usual statistical analysis of spin glass theory, makes no reference to the specific dynamics of the system.

As we expect that quantum computers find NP-complete problems just as troubling as classical computers, the classical problems that are hard to approximate remain so even exploiting quantum dynamics. However, this leaves open the question of how hard intrinsically \emph{quantum} (QMA-complete) problems \cite{Kitaev:1999ve,Aharonov:2002p4066,Bravyi:2006p4315} might be to approximate, i.e. of the
precise statement of a quantum analog of the PCP theorem. One possibility derives from the usual intuition of quantum statistical mechanics that a finite energy density
corresponds to a finite temperature which renders quantum mechanics essentially classical. If this intuition governs then we would expect to be able to efficiently represent
approximate finite energy density states for QMA-complete problems on a classical computer and thus the approximation problem would ``only'' be in NP and thus NP-complete.
The other, more interesting, possibility is that the above is not true and that there exist quantum Hamiltonians whose low energy \emph{density} states are both hard to find
and impossible to represent with an efficient classical description. Here even the approximation problem would be in QMA. Of course these possibilities are not
exclusive---possibly there is a ``phase transition'' as a function of the energy density where the problem goes from NP to QMA as the energy density is lowered.

In this paper we take an approach to this question which fits naturally within the quantum statistical mechanics of random systems
wherein, unlike in the Computer Science setting, we do not attempt to make statements about all possible instances of a problem
but instead pick a natural measure on the space of problems and try to investigate the properties of instances which are typical
with respect to it. Specifically, we consider the approximation problem for the low energy states of a canonical QMA-complete problem: $k$-body quantum satisfiability ($k$-QSAT).
As in previous work \cite{Laumann:2010fk,Laumann:2010kx,Bravyi:2009p7817,Ambainis:2009fv}, we consider a uniform random ensemble of $k$-QSAT instances with $M$ clauses, $N$ qubits and clause density $\alpha = M/N$ held fixed in the thermodynamic limit.

\begin{figure}[htbp]
	\centering
		\includegraphics[width=\columnwidth]{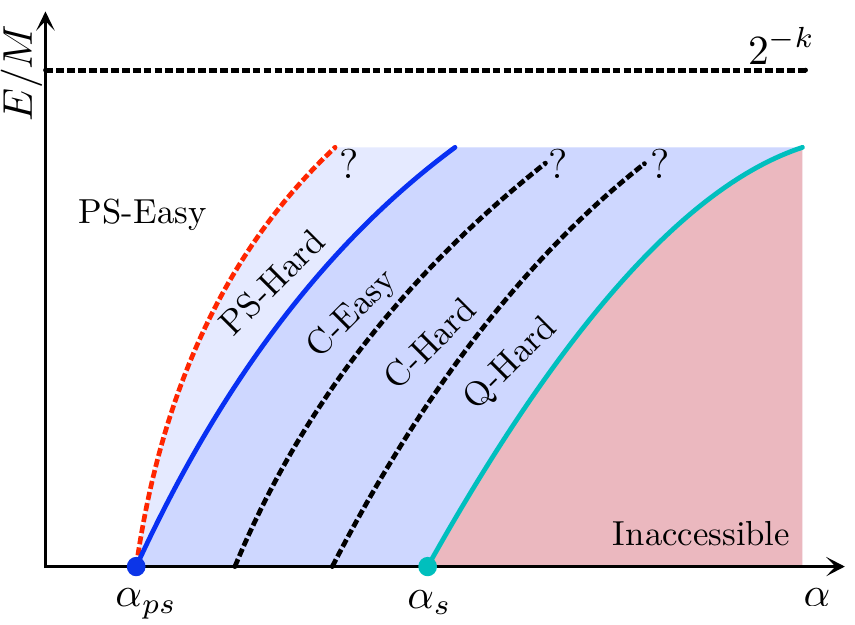}
	\caption{Heuristic `hardness' phase diagram from approximating random QSAT. PS regions have product state witnesses, C have classical non-product and Q have only quantum witnesses. ``Easy" and ``Hard" refer to conjectural algorithm-dependent hardness transitions for finding such witnesses.}
	\label{fig:cartoon-pd}
\end{figure}

The current state of knowledge and our general program with regard to this ensemble is summarized in Fig.~\ref{fig:cartoon-pd} where we
consider a ``phase diagram'' in a plane labeled by $\alpha$ and the energy per clause. It is known that  typical instances are satisfiable (i.e. have strictly zero energy ground states) by product states  at low density,  $\alpha < \alpha_{ps}$,
satisfiable by entangled states for intermediate densities $\alpha_{ps} < \alpha < \alpha_{s}$, and strictly unsatisfiable for
high densities $\alpha > \alpha_s$. For $k=3$, of maximum interest in this paper, $\alpha_{ps} = 0.91\dots$ and
$\alpha_{s} = 1.0 \pm 0.06$ \cite{Laumann:2010kx,Ambainis:2009fv}.\footnote{Current rigorous arguments provide $\alpha_{ps} = 0.91\dots$ and that $\alpha_s < 3.59$ for $k=3$ body interactions. The estimate of $\alpha_s = 1.0\pm 0.06$ comes from small-size numerics. The separation of $\alpha_{ps}$ and $\alpha_s$ and the accompanying entangled-SAT phase can be shown rigorously for $k\ge 12$.}. The questions we
would like to answer are denoted by a set of conjectured ``phases'' in the Fig.~1. Specifically, these are first to establish the
boundaries of the regions where
\begin{itemize}
  \item it is possible to use product states to achieve the required energy densities
  \item it is possible to use non-product classical states to achieve the required energy densities
  \item it is essential to use quantum states to achieve the required energy densities
  \item it is not possible to achieve the specified energy densities
\end{itemize}
By ``classical states'' we mean states for which energies can be evaluated in a time poly$(N)$ by a classical computer. Additionally, there is
the question of how long it takes to find such states. In principle for each category of states there can be algorithm
dependent easy and hard regions. It is also a challenge to determine these boundaries. Needless to say, it is not known
{\it a priori} whether all of these regions actually exist (apart from the product state and inaccessible regions).
We do know though that the random ensemble exhibits a finite surface/volume ratio for subgraphs of its core; this `expansion'
property makes it resistant to simple divide-and-conquer approximation strategies that apply to finite dimensional graphs and thus leaves open the possibility of hard-to-approximate phases.

In this paper we take a step towards answering the above questions by attempting to find the boundaries for the product
state regions.
Working over the variational class of product states for QSAT defines a classical, continuous spin, energy functional on the same interaction graph. We dub this the PSAT model (see Eq.~\eqref{eq:psat_ham}).
The analysis of this model is the central technical goal of this paper.
By introducing a classical Gibbs ensemble with temperature $T=1/\beta$ onto PSAT, we can use statistical mechanical techniques adapted from cavity analysis \cite{Mezard:2009vn,Mezard:2001p84} to probe the structure of the finite energy density landscape of this classical model. 
We note that cavity analysis of the associated classical model is similar in spirit to the cavity analysis of the coherent state representation of the AKLT spin glass \cite{Laumann:2008zl,Laumann:2010uq}, but the quantum-classical mapping involved here is approximate.
We will extract complementary information regarding the low energy structure by direct numerical optimization of finite-size instances using both a local greedy quench algorithm and simulated annealing.

Using these approaches, our aims are two-fold.
First, we extract the ground state energy density of PSAT as a function of density $\alpha$, as this defines the energy density below which QSAT has no product state approximation.
Second, we accumulate evidence for the existence of a classically `hard' phase at energy densities approaching the ground state boundary from above.
This phase exhibits metastability in the quench dynamics at finite size and a `dynamical replica symmetry breaking' instability in our cavity analysis \cite{Mezard:2009vn}.
In passing, we will provide a characterization of the low and high temperature thermodynamics of PSAT.

Finally we note that our work here is complementary to more rigorous partial results on the quantum PCP problem. 
The development of a quantum gap amplification procedure \cite{Aharonov:2008p6365} provided a quantum analogue of the central step in Dinur's proof of the classical result \cite{Dinur:2007bx}. 
However, the quantum no-cloning theorem has stymied attempts to complete a quantum PCP proof along those lines. 
Rather, the field has turned to constructing classical approximation strategies and classical approximate witnesses to restricted classes of Hamiltonians in an attempt to constrain possible quantum PCPs. 
For example, a now classic result is that \emph{commuting} 2-local Hamiltonians have only short-range entangled ground states \cite{Bravyi:2005ux} which possess efficient classical representations and no topological order. 
Several workers have extended these considerations to finite energy density witnesses to some classes of $k>2$-local commuting Hamiltonians \cite{Aharonov:2011bs,Aharonov:2013wv,Hastings:2012tb,Freedman:2013ue} under various restrictions.
Of particular relevance to us is the result that commuting Hamiltonians wherein a small fraction of the terms can be dropped to render the interaction complex 1-localizable possess classically representable states of low energy density \cite{Hastings:2012tb,Freedman:2013ue}. We note that the random ensemble we study in this work is indeed 1-localizable but not commuting.
Ground states constructed for commuting Hamiltonians may be perturbatively extended to finite energy density states for almost-commuting Hamiltonians \cite{Arad:2011wx}. 
Finally, for general (non-commuting) quantum Hamiltonians, the rigorous results have relied on constructing guarantees that product states \cite{Gharibian:2011p10667} or coarse grained product states (in the 2-local case) \cite{Brandao:2013uy} exist which provide good approximations to the ground state energy under certain geometric constraints on the interaction graph. 
Of these results, we note that low degree expanders as we study in this paper are precisely in the class that evades existing efficient approximation guarantees.

In the balance of the paper we do the following. 
We describe in the next section the random QSAT model, its relationship to the classical PSAT model and various pieces of useful terminology. Section \ref{sec:temp_expansion} discusses various limiting expressions for the PSAT thermodynamic quantities. Some details of these calculation can be found in the appendices. These analytic results serve as a useful test of the numerics.
Next, we present results from our greedy local quench, simulated annealing and exact diagonalization algorithms in Section \ref{sec:fin_numerics}. 
In Section \ref{sec:cavity_approach} we analyze the product state phase with the cavity method where we find some evidence for a dynamical instability. 
Finally, we give concluding remarks in Section \ref{sec:conclusion}. 
As an aid to the reader, we collect several results on the Haar measure on the space $\mathbb{CP}^{n-1}$ in Appendix \ref{sec:uniform_measure} and hypercores in Appendix \ref{sec:hypercore_datum}.


\section{Model} 
\label{sec:model}

The QSAT Hamiltonian on $N$ qubits is given by
\begin{align}
	\tilde{H} = \sum_m \Pi^m
\end{align}
where $\Pi^m$ is a $k$-local rank 1 projector associated with the (hyper-)edge $m = (m_1, m_2,\cdots, m_k)$ of an interaction graph $G$. $\Pi^m$ projects the state of $k$ qubits onto one `direction' in their $2^k$-dimensional joint Hilbert space, exacting on energy cost given by the size of the projection.
Such a Hamiltonian has a zero energy state $\ket{\Psi}$ if and only if $\ket{\Psi}$ is simultaneously annihilated by all of the projectors $\Pi^m$.
In this case, we say that $\tilde{H}$ is \emph{satisfiable} and that each of the \emph{clauses} $\Pi^m$ is satisfied by the state $\ket{\Psi}$.
We note that deciding whether a given QSAT Hamiltonian $\tilde{H}$ is satisfiable is QMA$_1$-complete for $k\ge 3$ \cite{Bravyi:2006p4315,Gosset:2013a}, subject to the technical restriction that the ground state energy of $\tilde{H}$, if non-zero, is larger than a polynomially small promise gap $\delta = 1/\rm{poly}(N)$.
As we expect that the energetics of QSAT in the ensembles we consider are extensive, the promise gap should be satisfied with high probability and we will not worry about it further.

The classical PSAT Hamiltonian on $N$ vector spins is given by the variational energy of the QSAT Hamiltonian $\tilde{H}$ over the class of qubit product states:
\begin{align}
	\label{eq:psat_ham}
	H(\set{\hat{n}_i}) &= \sum_m\bra{\set{\hat{n}_i}_{i \in \partial m}}\Pi^m\ket{\set{\hat{n}_i}_{i \in \partial m}} \\
	& = \sum_m|\sand{\phi^m}{\set{\hat{n}_i}_{i \in \partial m}}|^2
\end{align}
Here, $\ket{\hat{n}_i}$ is the spin-1/2 coherent state pointing in the direction of the unit vector $\hat{n}_i$ at site $i$ and $\ket{\phi^m}$ is the $k$-local state onto which the projector $\Pi^m$ projects. In complex coordinates, the Hamiltonian takes on the $k$-quadratic form
\begin{align}
	H = \sum_m |\phi^m_{a_1\cdots a_k} z_{m_1}^{a_1} \cdots z_{m_k}^{a_k}|^2
\end{align}
where the $a$ indices run over the two components of the spinor $z$. In this representation, the Hamiltonian is a simple polynomial, but the coordinates $z$ are constrained to be normalized $z^a (z^a)^* = 1$ and the local phase rotation $z^a \to e^{i \theta} z^a$  is redundant.
As usual, a convenient choice for the spinor $z$ associated to the direction $\hat{n}$ is
\begin{align}
	z = \begin{pmatrix}\cos(\theta/2) \\
						 \sin(\theta/2) e^{i \phi}\end{pmatrix}
\end{align}
where $\theta, \phi$ are polar coordinates for the direction $\hat{n}$.

Finally, let us define the random ensemble for $k$-P/QSAT in more detail.
There are two sources of randomness in the ensemble:
1. the discrete choice of interaction graph $G$ and
2. the continuous choice of projectors $\Pi^m$ associated to each edge.
The latter choice is particularly powerful: \emph{generic} choices of projectors reduce quantum satisfiability to a graph, rather than Hamiltonian, property \cite{Laumann:2010fk}.
More precisely, for fixed $G$, the dimension $R_G = |\ker H|$ of the satisfying subspace of $H$ is \emph{almost always} minimal with respect to the choice of projectors $\Pi^m$.
Thus, if $G$ can be frustrated by \emph{some} choice of projectors, it is frustrated for almost all such choices. We refer to this as the ``geometrization'' property.

Also of particular interest to us is the result that zero energy product states for generic QSAT instances exist if and only if the factor graph $G$ has `dimer coverings' between its variables and clauses which cover all of the clauses \cite{Laumann:2010kx}. We review this construction in some detail below (see Sec.~\ref{sub:low_temperature}).
Throughout this paper, we use the term \emph{generic} to refer to the continuous choice of projectors and \emph{random} to refer to the choice of the graph.

As for the discrete choice of random interaction graph $G$, we follow previous work and use the Erd\"{o}s-Renyi ensemble with clause density $\alpha$, in which each of the potential $\binom{N}{k}$ edges appears with probability $p=\alpha N / \binom{N}{k}$. One of the most important geometric features of such interaction graphs is the existence of a (hyper-)core above a critical $\alpha_{hc}$. The hypercore $G' \subset G$ is the maximal subgraph of $G$ such that every node has degree at least two.
One can show that all non-core clauses in $G-G'$ may be satisfied by local product states, whether or not $G'$ is satisfiable, thus all of the nonzero energetics are associated with the core.


\section{Low and high temperature expansions} 
\label{sec:temp_expansion}

We begin our study of the PSAT model by considering the leading order low and high temperature expansions for the classical Gibbs ensemble,
\begin{align}
	Z = \int \prod_i D\hat{n}_i\, e^{-\beta H(\set{\hat{n}_i})}.
\end{align}
These limits provide non-trivial consistency checks for the more sophisticated numerical and cavity analyses in later sections.
As the model is classical, the low temperature expansion follows from equipartition, even in the presence of disorder averaging. However, one must be careful to take account of `zero modes'. A classical zero mode corresponds to a degenerate continuous manifold of ground states which therefore do not contribute to the specific heat. We work this out in Sec.~\ref{sub:low_temperature} below.

The disorder-averaged high temperature expansion is a straightforward exercise in combinatorics. We summarize the leading order details in Sec.~\ref{sub:high_temperature}.

\subsection{Low temperature} 
\label{sub:low_temperature}

Generic PSAT has zero energy satisfying states when the interaction graph $G$ possesses `dimer coverings': matchings of spins to clauses such that every clause is covered by exactly one dimer and no spin is used more than once \cite{Laumann:2010kx}.
Let us review the construction of these states.
The counting is easiest if we assume that the projectors are all of product form
\begin{align}
	\phi^m_{a_1\cdots a_k} = \phi^{m_1}_{a_1} \cdots \phi^{m_k}_{a_k}
\end{align}
where each $2$-spinor $\phi^{m}$ is individually normalized.
A zero energy product state may be associated to each dimer covering of the factor graph: if site $i$ is paired with clause $m$ by the dimer covering, we use site $i$ to satisfy clause $m$ by making it orthogonal to the relevant factor ($\epsilon^{ab}$ is the antisymmetric tensor)
\begin{align}
	z^a_i = \epsilon^{ab} \phi^{i}_b.
\end{align}
This leaves $N-M$ spinors completely unspecified -- these are zero modes of the system. Meanwhile, any deviation of the matched spinor at site $i$ leads to a quadratic energy cost for generic choices of the unspecified spinors, so that there are $M$ complex harmonic modes, or $2M$ real harmonic modes. Recalling that the ground state energy is $0$, equipartition produces
\begin{align}
	U \approx T M = T \alpha N.
\end{align}

This counting persists for generic projectors (not of product form).
To recap \cite{Laumann:2010kx}, let $\set{z_i}$ be a zero energy ground state of $H$. Since $H(\set{z_i}) = 0$, the $z_i$ must satisfy
\begin{align}
	\phi^m_{a_1\cdots a_k} z^{a_1}_{m_1} \cdots z^{a_k}_{m_k} = 0 \qquad \forall m=1\cdots M
\end{align}
By rotating the local bases, we can assume that the ground state $z_i = \begin{pmatrix}1\\0\end{pmatrix}$ for all sites $i$, in which case these equations simply reduce to $\phi^m_{0\cdots 0} = 0$.
Perturbing $z_i \to \begin{pmatrix}1\\\delta w_i\end{pmatrix}$, we find (to leading order)
\begin{align}
	\label{eq:linearized_psat}
	\sum_{j=1}^k \phi^m_{0\cdots1\cdots0} \delta w_{m_j} = 0
\end{align}
The dimension of the kernel of this set of $M$ linear equations in $N$ unknowns $\delta w_i$ gives the number of (complex) zero modes. For factor graphs $G$ with dimer coverings, this linear map is generically surjective so the dimensional of the kernel is $N-M$ and we recover the counting from the product projector case.

For graphs that are not product satisfiable, the ground state energy is non-zero on the hypercore of the graph $G$ and we expect no zero modes associated with the core spins.
Nonetheless, there may still be zero modes on the non-core spins. An equipartition based estimate of the low temperature energy density in the non-product satisfiable phase looks like:
\begin{align}
	\label{eq:U_lowT_highalpha}
	U&\approx M \epsilon_0 + T(N_{c} + (M-M_{c}))
\end{align}
where $\epsilon_0$ is the ground state energy per clause ($\epsilon_0>0$), $N_{c}$ is the number of spins on the core, $M_c$ is the number of clauses on the core. For random interaction graphs $G$, the geometric quantities $N_c(\alpha)$ and $M_c(\alpha)$ are known \cite{Mezard:2003p5977} and we have quoted representations for them in Appendix~\ref{sec:hypercore_datum}.

Of course, this estimate only applies at sufficiently low temperatures that all of the non-zero modes indeed look harmonic.


\subsection{High temperature} 
\label{sub:high_temperature}

The high temperature expansion for the disorder averaged free energy is straightforward:
\begin{align}
	\overline{F} &= -\f{1}{\beta} \overline{\log Z(0) } + \overline{\avg{H}}-\f{\beta}{2!}\overline{\avg{H^2}-\avg{H}^2}  \nonumber\\
&	+ \f{\beta^2}{3!}\overline{\avg{H^3} - 3\avg{H}\avg{H^2} + 2\avg{H}^3}  \nonumber \\
&	- \f{\beta^{3}}{4!} \overline{ \avg{H^{4}} - 4\avg{H^{3} }\avg{H} - 3\avg{H^{2} }^{2} + 12\avg{H^{2} } \avg{H}^{2} - 6\avg{H}^{4} } \nonumber \\
&	+\dots \nonumber\\
	&= -\f{1}{\beta} C^{(0)} + C^{(1)} - \f{\beta}{2!} C^{(2)} + \f{\beta^2}{3!} C^{(3)} - \f{\beta^3}{4!} C^{4} \cdots
\end{align}
where  $\avg{\cdot}$ denotes the infinite temperature average and $\overline{\cdot}$ denotes the disorder average.
Normalizing the volume of the spinor space to $4\pi$, we have
\begin{align}
	\log Z(0) = N \log 4\pi
\end{align}
independent of disorder realization. 

Both the thermal and disorder correlation functions which arise in evaluating the high temperature expansion are with respect to normalized vectors on complex spherical spaces. These satisfy a variant of Wick's theorem which we derive in Appendix~\ref{sec:uniform_measure}. For example, to evaluate the disorder averaged infinite temperature energy:
\begin{align*}
	\overline{\avg{H}_0} &= \sum_m\overline{\avg{E^m}} \\
	& = M\overline{\avg{|\phi_{a_1\cdots a_k} z_1^{a_1}\cdots z_k^{a_k}|^2}} \nonumber\\
	&= M \overline{\phi_A\phi^*_B} \avg{z^{a_1}_1(z^{b_1}_1)^*}\cdots \avg{z^{a_k}_1(z^{b_k}_1)^*} \nonumber\\
	&= M \f{\delta_{AB}}{2^k} \f{\delta^{a_1b_1}}{2}\cdots\f{\delta^{a_kb_k}}{2}\nonumber\\
	&= \f{M}{2^k}
\end{align*}

In evaluating the higher order cumulants, two properties are especially important. First, as usual in the high temperature expansion, disconnected clusters vanish. Thus, the second order cumulant is
\begin{align}
	C^{(2)} &= \sum_{m,n} \overline{\avg{E^m E^n} - \avg{E^m}\avg{E^n}} \nonumber \\
	&= M (c^{(2)}_0 + \alpha k^2 c^{(2)}_1)
\end{align}
where the constants
\begin{align}
	c^{(2)}_0 &= \overline{\avg{(E^0)^2}-\avg{E^0}^2}\\
	c^{(2)}_1 &= \overline{\avg{E^0E^1}-\avg{E^0}\avg{E^1}}
\end{align}
correspond to the single clause energy fluctuations and the overlapping clause energy fluctuations and $\alpha k^2$ is the expected number of clauses intersecting a given clause $m$ in one spin. 

Second, the disorder average of any power of a single clause energy function produces a number with no thermal fluctuations.
\begin{align}
	\overline{(E^0(z))^n} = \f{n!}{2^k(2^k+1)\cdots(2^k+n-1)}
\end{align} 
Thus, in any term in the cumulant expansion where all copies of a clause lie in the same thermal average, we may disorder average first and extract the constant. This implies that any cumulant in which a particular clause $E^m$ arises only once vanishes, as `constant' random variables always lead to vanishing joint cumulants. For example, this tells us immediately that $c^{(2)}_1 = 0$.

\begin{table}
	\begin{center}
	\renewcommand{\arraystretch}{1.5}
	\begin{tabular}{c|c|c|r@{$\times$}l|c}
	Order 	& Symbol &	Cumulant	  & \multicolumn{2}{|c|}{Value} & Geometry \\
	\hline\hline
	1	&	$c^{(1)}_0$ & $\overline{\avg{E^0}_c}$	& $1$& $2^{-3}$ &$M$ \\ 
	\hline
	2	& 	$c^{(2)}_0$ & $\overline{\avg{(E^0)^2}_c}$	& $7/9$& $2^{-6}$ & $M$ \\ 
	\hline
	3	&	$c^{(3)}_0$ &$\overline{\avg{(E^0)^3}_c}$	& $14/15$ & $2^{-9}$ & $M$ \\ 
	\hline
	4	&	$c^{(4)}_0$ &$\overline{\avg{(E^0)^4}_c}$	& $5078/4455$ & $2^{-12}$ &$M$ \\
		&	$c^{(4)}_1$ &$\overline{\avg{(E^0)^2(E^1)^2}}_c$		& $-2/243$ & $2^{-12}$  & $3 M \alpha k^2$ \\
	\end{tabular}
	\end{center}
	\caption{The non-vanishing terms in the high temperature expansion up to order 4 explicitly for $k=3$. The $\avg{\cdot}_c$ indicates the symmetrized joint cumulant of the clause energy variables. The geometric factor indicates the expected number of such terms that contribute to the free energy in the random graph. We break out the factors of $2^{-kn}$ ($n$ being the order) in the value as we expect the terms to scale as such. }
	\label{tab:hightemp}
\end{table}

Using these rules, we find that the only terms that contribute to the high temperature expansion to 4th order are those given in Table~\ref{tab:hightemp}. We have included the explicit evaluation of those terms for $k=3$.

To summarize, the thermodynamic densities to fourth order are
\begin{align}
	F/N &= -\frac{1}{\beta} \log 4\pi + \alpha / 2^k - \f{\beta \alpha}{2}c^{(2)}_0 + \f{\beta^{2}\alpha}{3 !}  c^{(3)}_{0} \nonumber\\
	& - \f{\beta^{3}\alpha}{4!}\left( c^{(4)}_{0} + 3\alpha k^{2} c_{1}^{(4)} \right)\label{eqn:high_temp1}\\
	S/N &= \log 4\pi - \beta^2 \f{\alpha}{2}c^{(2)}_0 + \beta^{3} \f{\alpha}{3} c_{0}^{(3)} \nonumber \\
	& -\beta^{4} \f{10 \alpha}{4!}  \left( c_{0}^{(4)} + 3\alpha k^{2} c_{1}^{(4)}\right) \\
	U/N &= \alpha\bigg[\f{1}{2^k} - \beta c^{(2)}_0 + \f{\beta^{2}}{2} c_{0}^{(3)} \nonumber \\
	& - \f{\beta^{3} }{3!} \left( c_{0}^{(4)} + 3\alpha k^{2} c_{1}^{(4)}\right) \bigg]. \label{eqn:high_temp3}
\end{align}



\section{Finite-size numerics} 
\label{sec:fin_numerics}

The general considerations regarding zero energy states and zero modes provide a surprisingly detailed picture of the satisfiable phase of PSAT, including its low energy density of states in both the satisfiable and non-satisfiable phases.
However, these arguments fail to provide any estimate of the ground state energy density when it becomes finite, nor do they help to understand the landscape of the search problem for an algorithm attempting to optimize a particular instance.
In this section, we turn to several complementary finite-size numerical studies in order to pin down the PSAT ground state energy density and compare it to the QSAT ground state energy density.

\begin{figure}[t]
	\centering
		\includegraphics[width=\columnwidth]{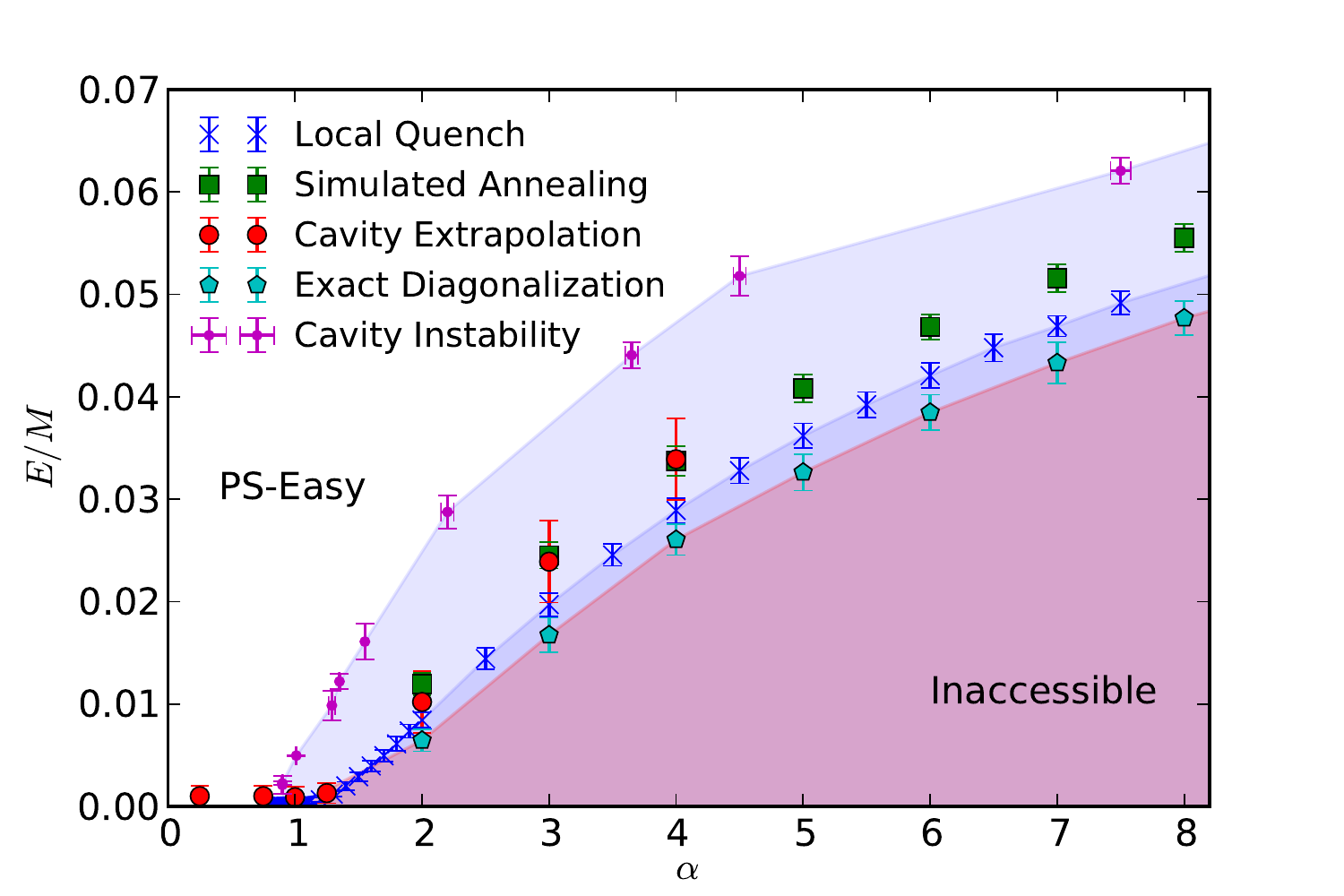}
	\caption{Finite energy density phase diagram for $3$-QSAT extracted from various numerical methods. Dark blue is QSAT inaccessible region, as estimated by exact diagonalization for size $N=20$. Medium blue region is PSAT inaccessible region, as estimated by local quench for size $N=100$. The light blue region indicates where the cavity analysis of PSAT exhibits a dynamical instability. More details regarding the numerical techniques and averaging are in the main text.}
	\label{fig:energy-bdy}
\end{figure}

\subsection{Greedy local quench} 
\label{sub:greedy_local_quench}

The most successful numerical line of attack on the PSAT ground state energy density turns out also to be the simplest: the greedy local quench.
The algorithm proceeds as follows:
Given a PSAT instance $H$, generate a random initial spin configuration $z_i(0)$ and then iteratively sweep over the sites $i$ in some fixed order. At each step, update the local spin to point in the direction which optimizes the energy of the neighboring clauses:
\begin{align}
	z_i(t+1) = \argmin_{z_i} \sum_{m \in \partial i} E^m(z_i,\set{z_j(t)}_{j\in \partial m - i}).
\end{align}
Here, the integer $t$ keeps track of the number of sweeps in the algorithm.
The calculation of the local minimizer is straightforward.
One simply calculates an effective field $\vec{h}_m$ on site $i$ due to each clause $m$ and then orients the spinor $z_i$ in the direction of the net field $\sum_m \vec{h}_m$.
As the energy function decreases with every step of the quench, the algorithm is guaranteed to converge toward a local minimum of the energy.
We terminate the quench when the relative change in the energy $\delta E/E$ after a full sweep across the $N$ sites is less than a fixed threshold ($10^{-4}$ in all data shown).
In applying the quench to the random ensemble, we have generated several hundred instances of size $N=100$ at varying clause density $\alpha$, stripped them to their hypercores (as all clauses not in the hypercore can be satisfied independent of the state of the core) and run the quench on each core with ten different initial conditions, taken the lowest resulting energy.

\begin{figure}[t]
	\centering
		\includegraphics[width=\columnwidth]{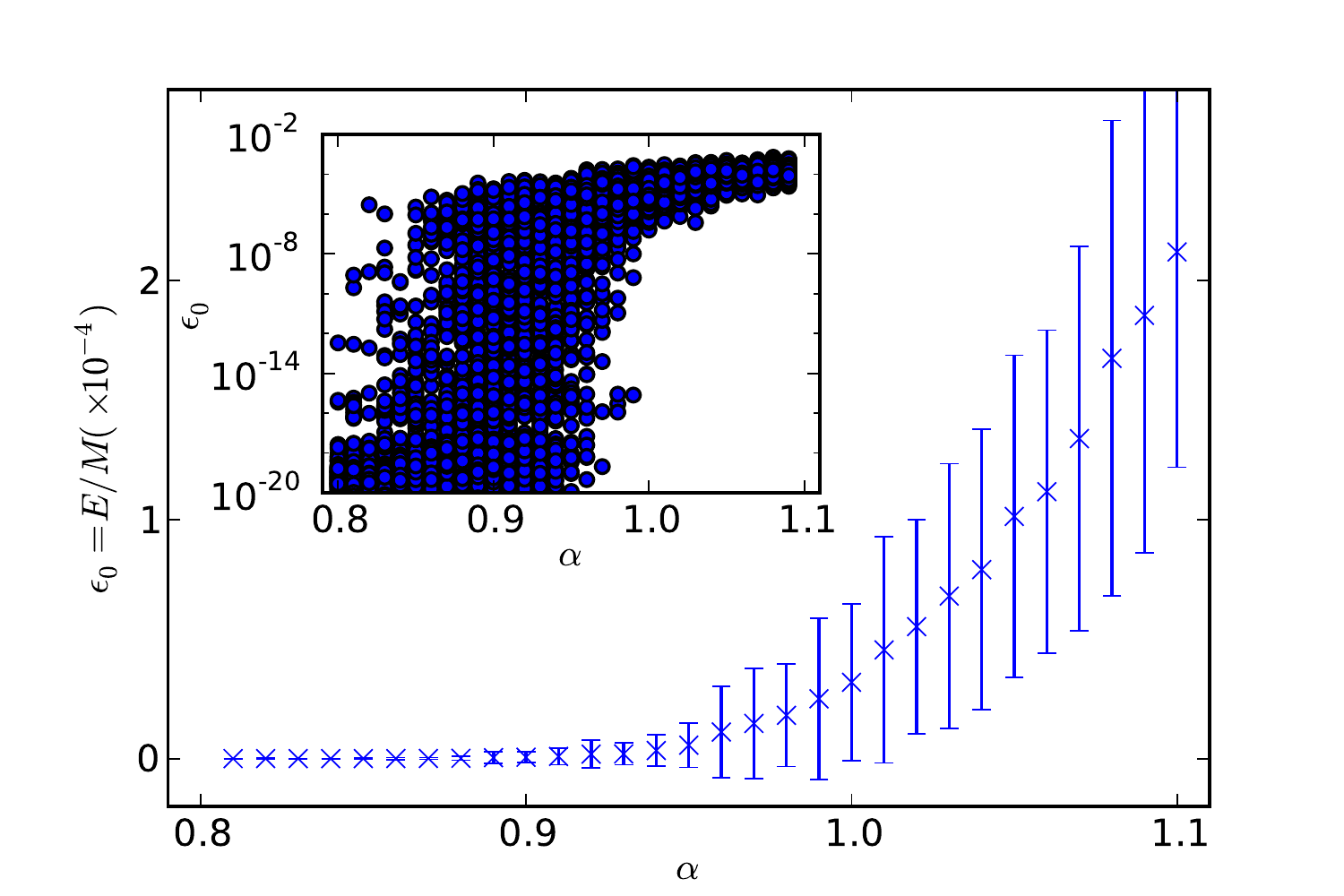}
	\caption{Product state ground state energy found by greedy local quench near critical region $\alpha_{ps}\approx 0.92$ at finite size $N=100$. Error bars indicate instance-to-instance fluctuations over $N_{samp}\approx 3-400$ instances per density $alpha$.
	(inset) Semilogarithmic scatter plot of same data.}
	\label{fig:quench-logscatter}
\end{figure}

The greedy quench does an excellent job of finding low energy product states in the random $3$-PSAT ensemble, as shown in Figure~\ref{fig:energy-bdy}.
The quench readily finds states with energies equal to zero to machine precision below $\alpha_{ps}= 0.9 \pm 0.06$ (see width of scatter region in Fig.~\ref{fig:quench-logscatter} inset).
Above this threshold, the found energies begin to grow from zero, in agreement with the rigorous expectation that the ground state energy is no longer precisely zero.
We note that the observed error in $\alpha_{ps}$ is consistent with the fluctuations in the core density due to finite-size $N=100$, which we can roughly estimate $N_c/M_c = 1 \pm \f{0.7}{\sqrt{N}}$ by noting that $N_c/N = M_c/N = 0.6$ in the thermodynamic limit and assuming that the choice of clauses and sites remaining in the core are selected by an independent random process from the initial random graph with size $N$. 

The observed energy density for $\alpha > \alpha_{ps}$ is both quantitatively very small -- even at $\alpha = 8$, the found energies of $0.05$ per clause are significantly below the naive high density estimate of $1/8 = 0.125$ per clause -- and the critical turn on for $\alpha$ near $\alpha_{ps}$ appears very soft (see inset, Fig.~\ref{fig:energy-bdy}). The simplest rigorous upper bound for the large $\alpha$ energy density is to satisfy the first $\alpha_{ps}N$ of the clauses and then upper bound the energy of the remaining clauses by $1/8$. This leads to $E/M\le (1-\alpha_{ps}/\alpha)/ 8$.
 
A better estimate at large clause density can be made by noting that every spin $i$ is subject to a mean exchange field due to the $d_i$ clauses attached to it, where the degree $d_i$ is a Poisson distributed random variable with mean $k \alpha$. 
With respect to an infinite temperature spin state, this exchange field is a sum of $d_i$ independent random variables so that it has $O(\sqrt{d_i})\sim O(\sqrt{k \alpha})$ fluctuations.
Thus, by aligning each local spin with its local exchange field, we expect to gain of order $O(\sqrt{k \alpha})$ energy per spin relative to the infinite temperature state. 
This produces a high density estimate of $E/M = 1/2^k - O(1/\sqrt{\alpha})$. 
We believe this essential scaling at large $\alpha$ is robust although the coefficient hiding in the $O$-notation is not so easy to produce -- the variance of the exchange field in an infinite temperature state is straightforward to calculate, but the process of tipping spins toward their local exchange field modifies the field on neighbors in a nontrivial manner.
In any event, we note that a fit (not shown) to this quadratic form at high clause density $\alpha > 2$ produces good agreement with an extracted coefficient of roughly $0.18 - 0.21/\sqrt{\alpha}$ (at $k=3$).

\begin{figure}[htbp]
	\centering
		\includegraphics[width=\columnwidth]{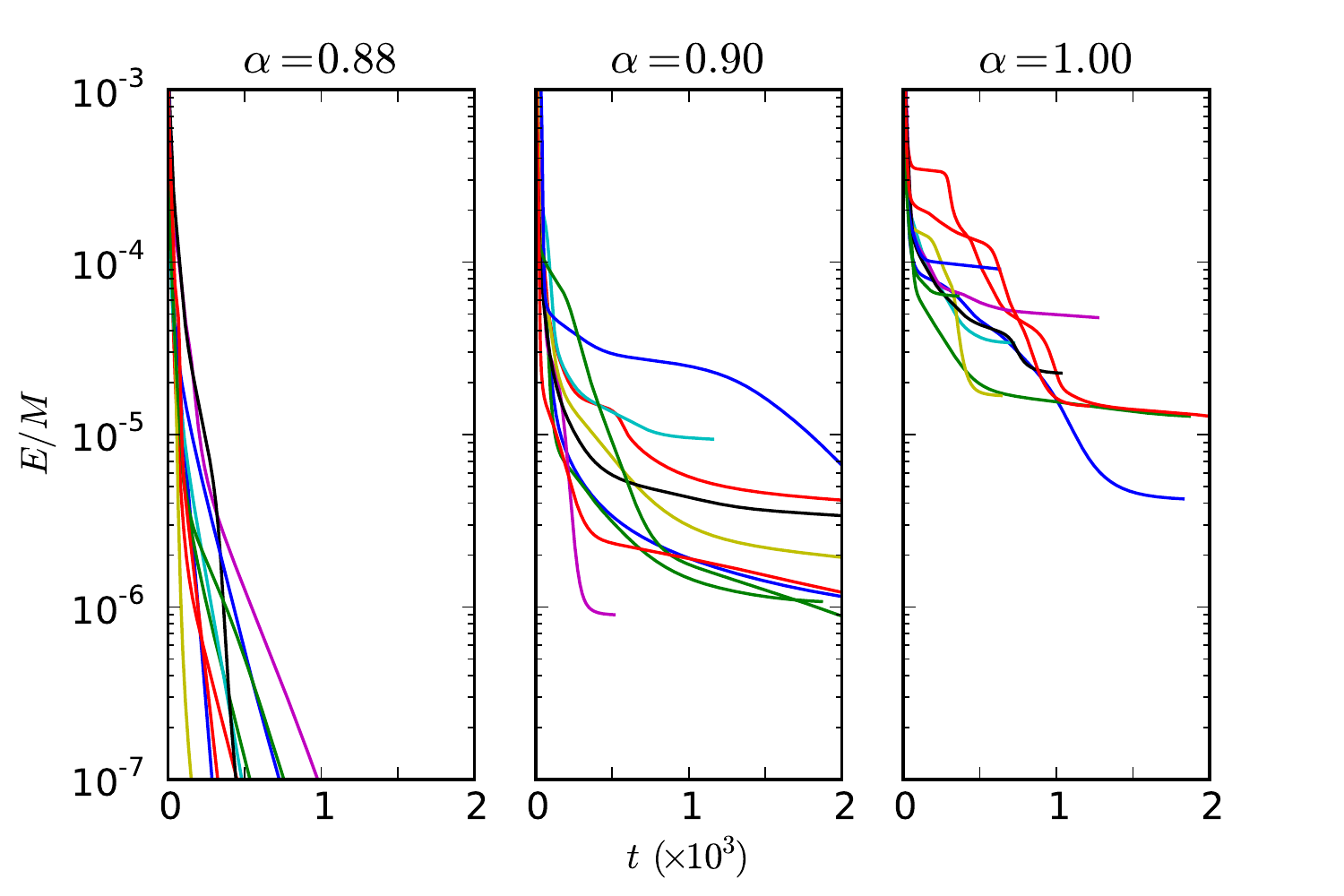}
	\caption{Typical energy traces as a function of time for 10 independent sweeps in a typical instance at each of $\alpha = 0.88, 0.90, 1.0$ ($N=100$). The curve in the left panel continue to fall exponentially (linearly on the log-scale) to machine precision ($10^{-16}$). }
	\label{fig:quench-timetrace}
\end{figure}

The dynamics of the local quench show mild evidence for the presence of metastable minima in the classical PSAT energy landscape above the transition region, as shown in the typical time traces shown in Fig.~\ref{fig:quench-timetrace}. For densities below $\alpha_{ps} = 0.91$, the typical traces decay continuously to very low values while at higher density, they exhibit a larger variance of found energies for different initial conditions and step-like metastable behavior in the individual trajectories.

\subsection{Simulated annealing} 
\label{sub:simulated_annealing}

As a second approach to finding low energy product states, we implemented a standard simulated annealing protocol to cool the PSAT Hamiltonian to low temperature on randomly generated instances. 
At each step of this procedure, a randomly sampled spin is perturbed by a small random step and the move is accepted following a Metropolis rule at temperature $T$. 
The temperature decays exponentially with a decay constant of $1.003$ per $10^4$ Monte Carlo sweeps, beginning with temperature $1$ and finishing at $T=10^{-8}$.

\begin{figure}[!t]
\centering
\includegraphics[width=0.9\linewidth]{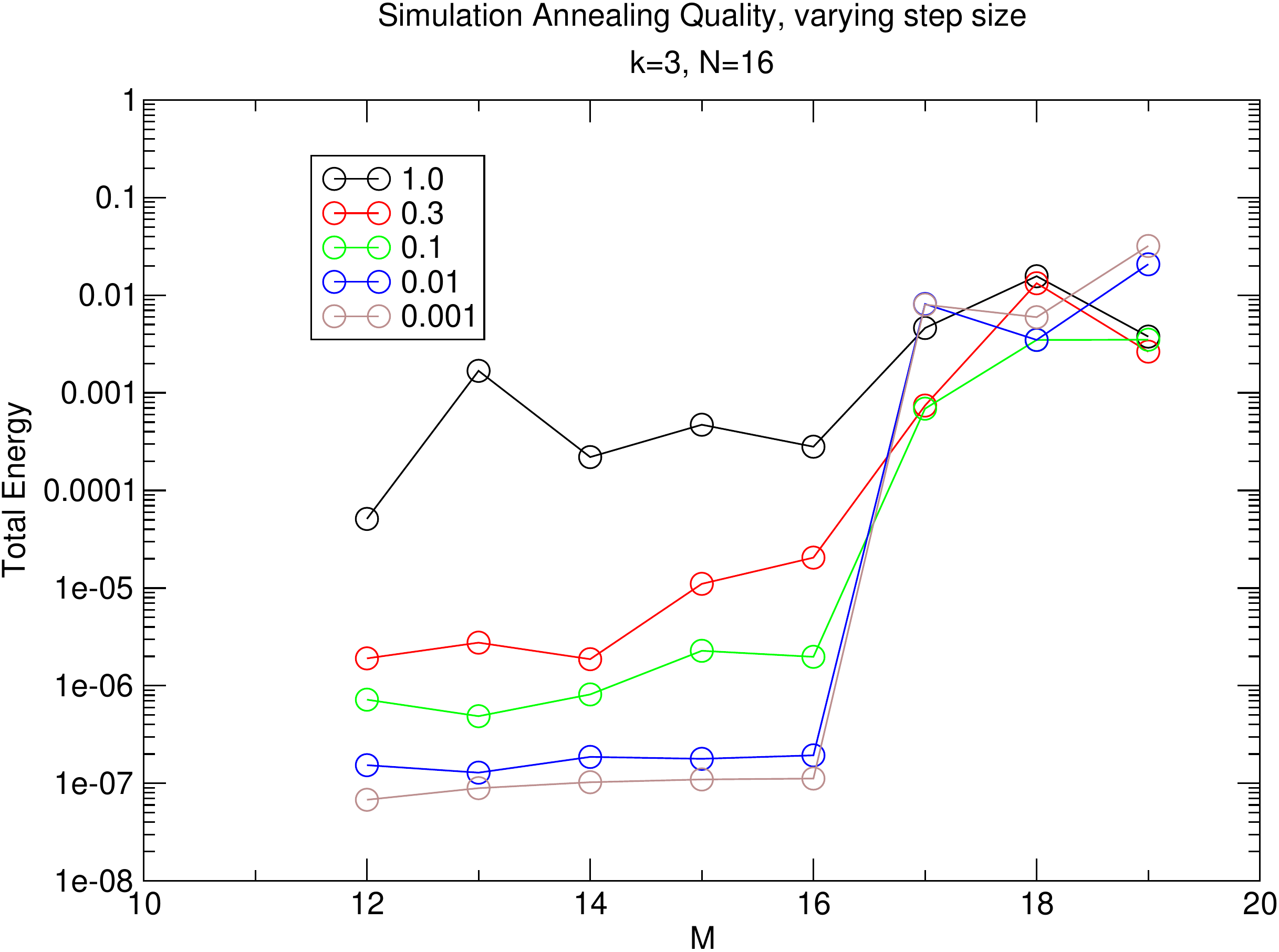}\\
\vspace{2mm}
\caption{\label{fig:anneal_steps}
Step size dependence of the best found PSAT energy in the simulated annealing algorithm.
The step size controls the maximum change of a single component of the wave function per update trial.
In the PRODSAT phase reducing the step size leads to a monotonically decreasing energy.
In the UNPRODSAT phase the behavior is irregular.
}
\end{figure}

The resulting disorder averaged energies $N=10-80$ qubits show extremely limited finite size scaling (not shown) so that we have included only the largest size data ($N=80$) in Fig.~\ref{fig:energy-bdy}. We note that simulated annealing fails to find energies as low as the local quench protocol but
agrees well with the low temperature extrapolation of the cavity results (see Sec.~\ref{sec:cavity_approach}).  While it is possible that the annealing results would improve by further optimizing the cooling schedule, we suspect that the timescale for this cooling may become very long in the UNSAT regime.
This is indicated by the transition in the behavior of the step size dependence of the simulated annealing results as the clause density is tuned from the SAT to UNSAT phase, see Fig.~\ref{fig:anneal_steps}.


\subsection{Exact diagonalization} 
\label{sub:exact_diagonalization}

Finally, we include exact diagonalization results for systems of up to $N=20$ qubits in Fig.~\ref{fig:energy-bdy}. These provide a numerical baseline for the QSAT ground state energy density in the UNSAT regime (striped symbols) and demonstrate its separation from the best found product states by local quenching. We note that the finite size scaling of these results looks quite converged for $k=3$ (not shown), as seen in previous studies as well \cite{Laumann:2010kx}, but that for larger $k$ there are still strong finite size effects for the achievable system sizes, as the graphs are quite unlike the thermodynamic random graph at these sizes. This is indeed the central reason that we limit the numerical investigations of this paper to the $k=3$ case.



\section{Cavity approach} 
\label{sec:cavity_approach}

In this section, we look at the PSAT model using so-called cavity techniques \cite{Mezard:2009vn}.
Our motivations for doing so are twofold. 
Firstly, while the previous finite size numerical work obtained low ground state energy estimates, these were limited to relatively small system sizes of at most $N=100$ variables.
With the cavity approach we can explore the thermodynamic limit where $N\rightarrow\infty$ directly.
Secondly, we also observed regions of the parameter space where the finite size algorithms began to slow down. In the local field quench, this slow-down coincided with what appeared to be metastable states, while in the simulated annealing results, the algorithm reached a barrier in finding lower energy ground states (see Figures \ref{fig:anneal_steps} and \ref{fig:quench-timetrace}).
However, these observations in a finite size system do not indicate whether these metastable states reflect true transitions in the thermodynamic structure of the configuration space.

The cavity approach allows us to detect the appearance of such states as $N\rightarrow \infty$. 
This is indicated by the appearance of a dynamical instability where the simplest (replica-symmetric) cavity ansatz breaks down \cite{Mezard:2001p84}.
From a algorithmic perspective, it is unclear whether such a boundary is truly  relevant, as indicated by the robust results of the local field quench.
However, in the spin glass lore, it is believed that such barriers ultimately limit local search procedures for discrete classical systems and we believe the discovery of similar instabilities in the continuous degree of freedom PSAT system represents an interesting avenue for future detailed exploration.

In the next section, we introduce the cavity approach appropriate to the PSAT problem and provide the thermodynamic quantities that can be obtained from it. After this, we discuss how the replica symmetry broken phase can be detected in the cavity approach.

\subsection{Cavity method}
\label{sec:cavity_method}

\begin{figure}[htbp]
	\centering
		\includegraphics[width=0.85\columnwidth]{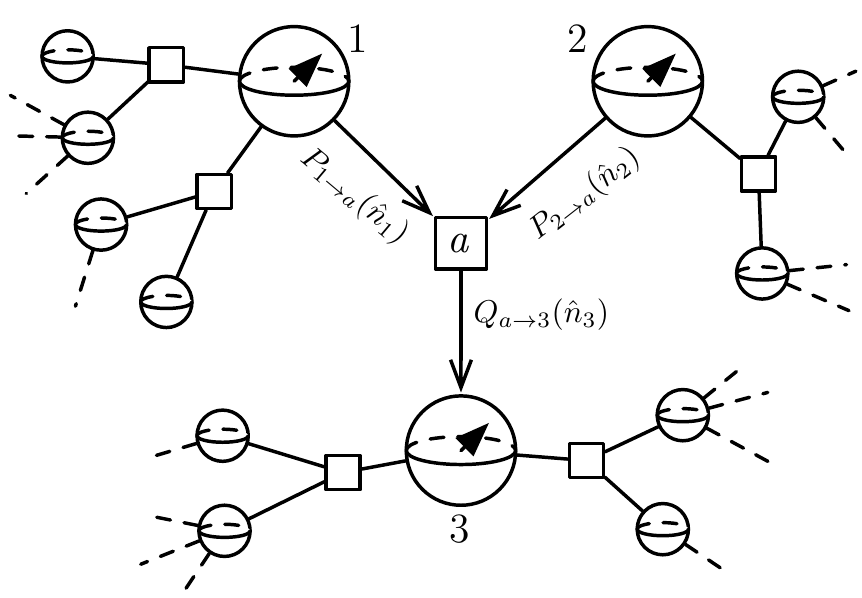}
	\caption{The definition of the cavity distributions on a locally tree-like interaction graph. Squares indicate clauses, labeled by letters, and spheres the spinor variables, labeled by integers.}
	\label{fig:bpgraph}
\end{figure}

The cavity approach to the classical PSAT model consists of studying the cavity distributions $P_{i\to m}(\hat{n}_i)$.
This is the marginal classical probability distributions for the unit vector $\hat{n}_i$ at site $i$, in the absence of the clause $m$.
The so-called belief propagation equations follow from assuming that the cavity distributions for the neighbors of the clause $m$ are independent when $m$ is removed.
Under this assumption, the cavity distributions as shown in Fig~\ref{fig:bpgraph} obey the equations
\begin{align}
	\label{eq:bp}
	Q_{m\to i}(\hat{n}_i) &= \f{1}{Z_{m\to i}} \int \prod_{j\in \partial m-i} \left(D\hat{n}_j  P_{j\to m}(\hat{n}_j)\right) e^{-\beta E^m(\set{\hat{n}_j}_{j\in\partial m})} \\
	P_{i\to m}(\hat{n}_i) &= \f{1}{Z_{i\to m}}\prod_{n\in \partial i - m} Q_{n\to i}(\hat{n}_i)
\end{align}
Here, $Q(\hat{n})$ is an intermediate cavity function that is useful primarily as a bookkeeping device.
For a $k$-local interaction graph with $M$ clauses, this provides $2Mk$ functional equations for $2Mk$ unknown cavity distributions $P(\hat{n})$ and $Q(\hat{n})$.

In the thermodynamic limit, we expect that the cavity distributions themselves are \emph{i.i.d} random variables with a self-averaging distribution $\mathbb{P}\left[P(\cdot)\right]$ and $\mathbb{P}\left[Q(\cdot)\right]$. Under this assumption, we obtain self-consistent equations for the functional distributions $\mathbb{P}\left[P(\cdot)\right]$ and $\mathbb{P}\left[Q(\cdot)\right]$:
\begin{widetext}
\begin{align}
	\label{eq:cavity-rs}
	\mathbb{P}\left[P(\cdot)\right] &= \mathbb{E}_{d,Q_n} \delta\left[ P(\cdot) - \frac{1}{Z}\prod_{n=1}^d Q_n(\cdot)\right] \\
	\mathbb{P}\left[Q(\cdot)\right] &= \mathbb{E}_{E^m, P_j}\delta\left[ Q(\cdot) - \frac{1}{Z}\int\prod_{j=1}^{k-1}D\hat{n}_j P_j(\hat{n}_j)e^{-\beta E^m(\set{\hat{n}_j}, \cdot)}\right]
\end{align}
\end{widetext}
These are the so-called replica-symmetric cavity equations for the PSAT model. Here, $\mathbb{E}$ stands for an expectation value with respect to the subscripted quantities, such as $d$, the onward degree of a spin, and $E^m$, the choice of projector on the clause. The normalization factors $Z$ implicitly depend on these random variables.

In general, the cavity equations \eqref{eq:cavity-rs} can not be solved exactly for the functional distributions $\mathbb{P}\left[P(\hat{n})\right]$, $\mathbb{P}\left[Q(\hat{n})\right]$.
While the relevant cavity equations are often analytically intractable even in models with discrete local degrees of freedom, such as in classical $k$-SAT, in PSAT, the cavity messages $P(\hat{n})$ and $Q(\hat{n})$ are general normalized probability distributions on the sphere, and the cavity distributions $\mathbb{P}\left[\cdot\right]$ are distributions on that functional space.
In the presence of sufficient symmetry, as, for example in rotationally invariant models \cite{Laumann:2010uq}, one can make progress in the symmetric phase analytically, but PSAT has no such rotational symmetry and we must resort to numerical methods throughout the phase diagram.

Thus, we follow established practice\cite{Mezard:2009vn,Zdeborova:2007lh, Zdeborova:2010ys, Zdeborova:2008mw} in studying cavity equations by representing the distributions, $\mathbb{P}\left[P(\cdot)\right]$ and $\mathbb{P}\left[Q(\cdot)\right]$, by a large \emph{population} of sampled messages $P(\cdot)$ and $Q(\cdot)$.
These probability distributions on the sphere in turn are represented by discretized probability vectors $P_i$, with a finite number $i=1\cdots N_{\textrm{disc}}$ of patches roughly equally sampled over the surface of the sphere.
By varying the discretization, we can check for convergence of thermodynamic observables to the continuous degree of freedom $N_{\textrm{disc}}\to\infty$ limit.

To solve the cavity equations \eqref{eq:cavity-rs}, we employ \emph{population dynamics}. We initialize the population of messages $P(\cdot)$ and $Q(\cdot)$ randomly and then proceed iteratively to use the belief propagation equations \eqref{eq:bp} to generate the new messages from the existing population.
As the population size $N_{pop}$ and total number of time steps $T$ approach infinity, we expect the resulting population to converge to a fixed point of the cavity equations \cite{Mezard:2009vn}. 
Of course, in practice, we take a fixed population size of $N_{pop} = 10^4$ and iterate until we observe statistical convergence of various moments.

Finally, with the converged messages in hand, one can estimate the various disorder averaged thermodynamic quantities such as the internal energy $U$ and the Bethe free energy, $F$.
\begin{widetext}
For example, the internal energy per clause follows directly from the expected value of the energy of a randomly added clause in the presence of $k$ sampled cavity messages.
\begin{align}
	U/M = \mathbb{E}_{E^a,P} \left[\f{1}{Z}\int D\hat{n}_1 D\hat{n}_2 D\hat{n}_3 E^a(\hat{n}_1,\hat{n}_2,\hat{n}_3) e^{-\beta E^a(\hat{n}_1,\hat{n}_2,\hat{n}_3) } P_{1\rightarrow a}(\hat{n}_{1}) P_{2\rightarrow a}(\hat{n}_{2}) P_{3\rightarrow a}(\hat{n}_{3} ) \right]
\end{align}
The expression for the Bethe free energy per variable $f = F/N$ is somewhat more complicated \cite{Mezard:2009vn}, as it follows from balancing the change in free energy of various graph surgery operations,
\begin{align}
	f (\alpha,\beta) = f_{v}(\alpha,\beta) + \alpha f_{a}(\alpha,\beta) - \alpha k f_{e}(\alpha,\beta)
\end{align}
where,
\begin{align}
	- \beta f_{v} &= \mathbb{E}_{\ell,Q} \log \left[ \int D\hat{n}\,Q_{1\rightarrow i}(\hat{n}) \dots Q_{\ell \rightarrow i}(\hat{n} ) \right] \nonumber \\
	- \beta f_{a} &= \mathbb{E}_{E^a,P} \log \left[ \int D\hat{n}_1 D\hat{n}_2 D\hat{n}_3 e^{-\beta E^a(\hat{n}_1,\hat{n}_2,\hat{n}_3) } P_{1\rightarrow a}(\hat{n}_{1}) P_{2\rightarrow a}(\hat{n}_{2}) P_{3\rightarrow a}(\hat{n}_{3} ) \right] \nonumber \\
	- \beta f_{e} & = \mathbb{E}_{P, Q } \log\left[ \int D\hat{n}\, P(\hat{n} ) Q(\hat{n}) \right].
\end{align}
\end{widetext}
These represent the change in the free energy from the addition of a single clause which has contributions coming from the variables added ($f_{v}$), the clause itself ($f_{a}$) and each edge connecting variable to edge ($f_{e}$).
Of course, with the internal energy $U$ and free energy $F$ in hand, it is straightforward to extract the entropy $S$ from the thermodynamic relation $F = U - TS$.

\subsection{Cavity thermodynamic results} 
\label{sub:cavity_thermodynamics}

\begin{figure}[t] 
	\includegraphics[width=.5\textwidth]{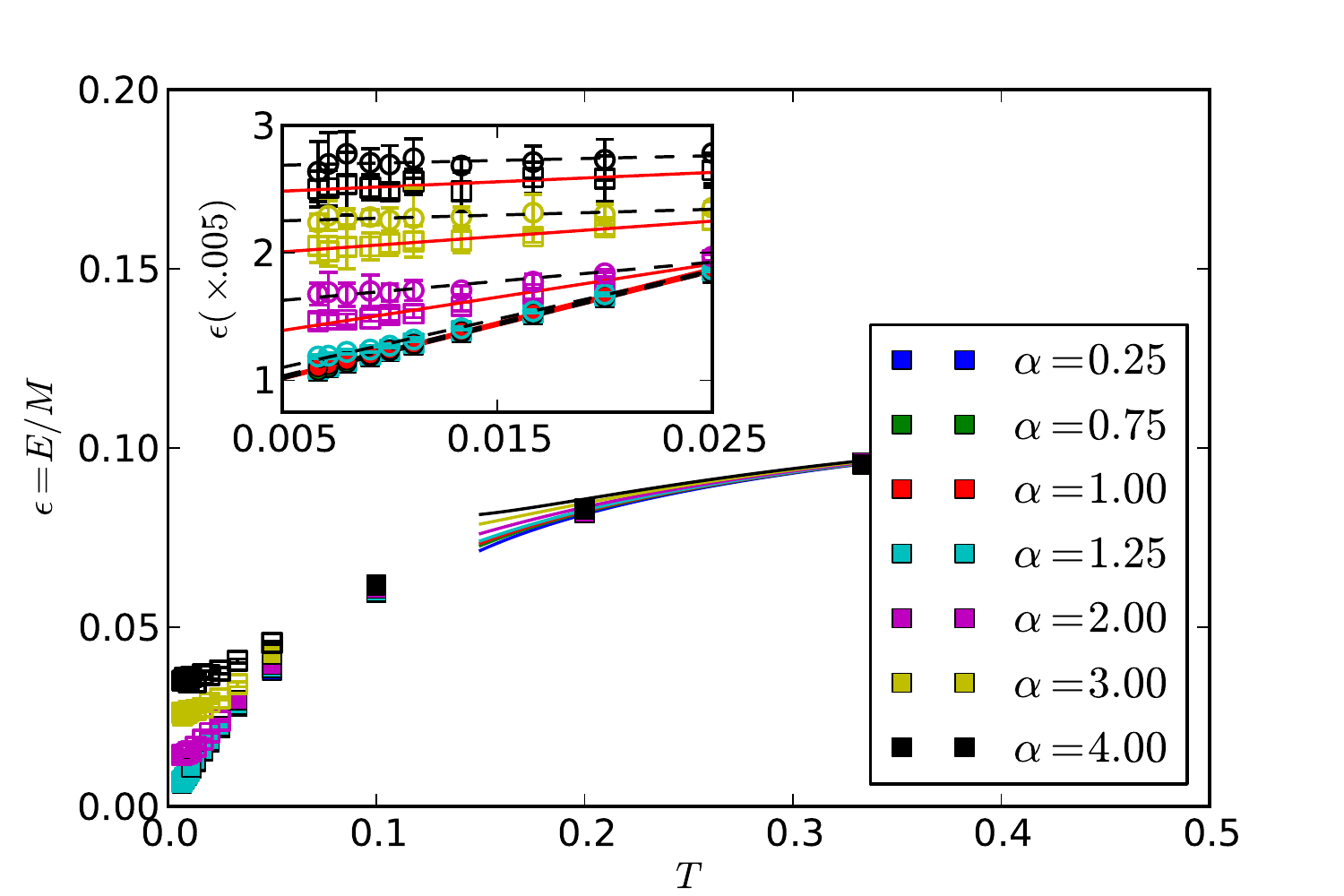}
	\caption{Cavity estimate of the internal energy of PSAT as function of temperature with $N_{disc} = 24$ states per Bloch sphere. Error bars reflect sampling error in calculating $E$ from the converged population. Solid lines indicate the high temperature expansion of the energy to 4th order. Solid (hollow) markers indicate data in the replica symmetric (broken) phase. 
	(inset) Zoomed in at low temperature of same data, including results for both  $N_{disc} = 24$ states ($\blacksquare$) and $N_{disc}=12$ states ($\bullet$). Solid (dashed) lines indicate $T$-linear fits to the $N_{disc}=24$ ($N_{disc}=12$) data. 
	 \label{fig:avgE}}
\end{figure}

\begin{figure}[t]
	\includegraphics[width=.5\textwidth]{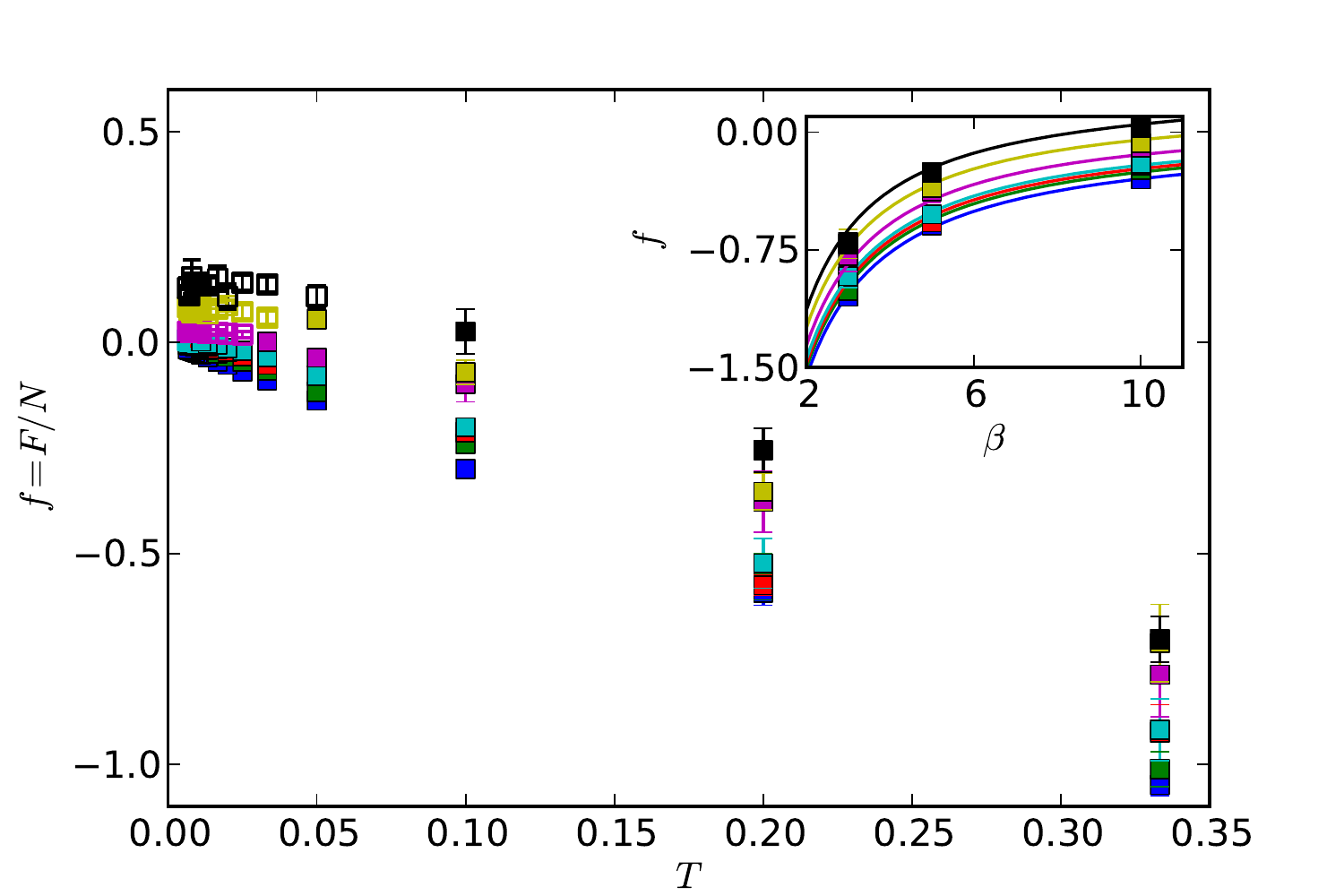}
	\caption{The free energy, $F/N$ as a function of $\alpha,T$ for qubits that have been discretized into 24 states. Point labeling is the same as in Figure \ref{fig:avgE} and likewise, hollow points indicate data collected ignoring the dynamical replica instability. (inset) As a function of $\beta$, we show the fit to the high temperature (small $\beta$) approximation for the free energy (cf. equations (\ref{eqn:high_temp1}) - (\ref{eqn:high_temp3}) ).  \label{fig:logZ}}
\end{figure}

We plot the energy density in Fig.~\ref{fig:avgE} and free energy density in Fig.~\ref{fig:logZ} at various temperatures, clause densities and  discretizations. 
The simplest results are at high temperature where the discretization error becomes small (not shown) and we expect detailed agreement with the high temperature expansion. 
Indeed, the expansion to 4th order, given by equations (\ref{eqn:high_temp1})-(\ref{eqn:high_temp3}), works extremely well for $\beta \lesssim 5-10$, as can be seen from the (no fit) agreement between the data and the solid lines in both the energy and free energy plots. 
We view this agreement as confirmation of both the high temperature expansion and the validity of the numerical cavity analysis.

At low temperature, we can fit the cavity results to the equipartition analysis of Sec.~\ref{sub:low_temperature}. 
In particular, we expect the energy to exhibit $T$-linear behavior with a zero-temperature intercept giving the ground state energy of the model and a slope corresponding to the specific heat. 
We have converged cavity populations down to temperatures of $T=1/500$ as shown in the inset to Fig.~\ref{fig:avgE}. 
The linear extrapolations provide an estimated ground state energy of roughly $10^{-4}$ for $\alpha = 0.25, 0.75, 1.00$ while for $\alpha= 1.25, 2, 3, 4$, the ground state energy is of order $10^{-2}$. 
These intercepts can be seen in Figure \ref{fig:energy-bdy}.  
We note that the statistical error bars in these estimates are quite large relative to the data found in the local quench numerics and, although they in principle exhibit no finite size effects, they do contain discretization error, which we expect to systematically raise the estimated energies and suppress the low temperature specific heat, as can be seen in the inset to Fig.~\ref{fig:avgE}.
Also, as we will discuss in more detail below, the low temperature-high density cavity calculation exhibits a dynamical instability below which we should hesitate to believe the quantitative results. 
Nonetheless, the ground state energies predicted by the cavity calculation at $N_{disc}=24$ are consistent with those found by simulated annealing and, at low density, with the zero expected for $\alpha < \alpha_{ps} = 0.91$.


\subsection{Dynamic instability} 
\label{sub:dynamic_instability}

In the context of discrete classical systems, the cavity equations \eqref{eq:cavity-rs} exhibit a well studied collection of instabilities that are believed to indicate the proliferation of solutions to the underlying belief propagation (BP) equations \eqref{eq:bp} on large instances \cite{Krzakala:2007p8722,Mezard:2009vn}. 
In turn, each of the many, extensively distinct solutions to the BP equations has been argued to correspond to a distinct thermodynamic pure state in the Gibbs ensemble. 
In the replica treatment of spin glass models, these instabilities are associated with replica symmetry breaking, and this language has been largely imported into the cavity formalism. 
From an algorithmic point of view, replica symmetry breaking indicates the appearance of metastable states that are `thermodynamically separated'. 
Such metastable configurations are believed to trap local search and annealing algorithms, frustrating the approach to lower temperatures -- a behavior that we observed in our finite-size numerics in the high clause density regime.

\begin{figure} 
	\includegraphics[width=.54\textwidth]{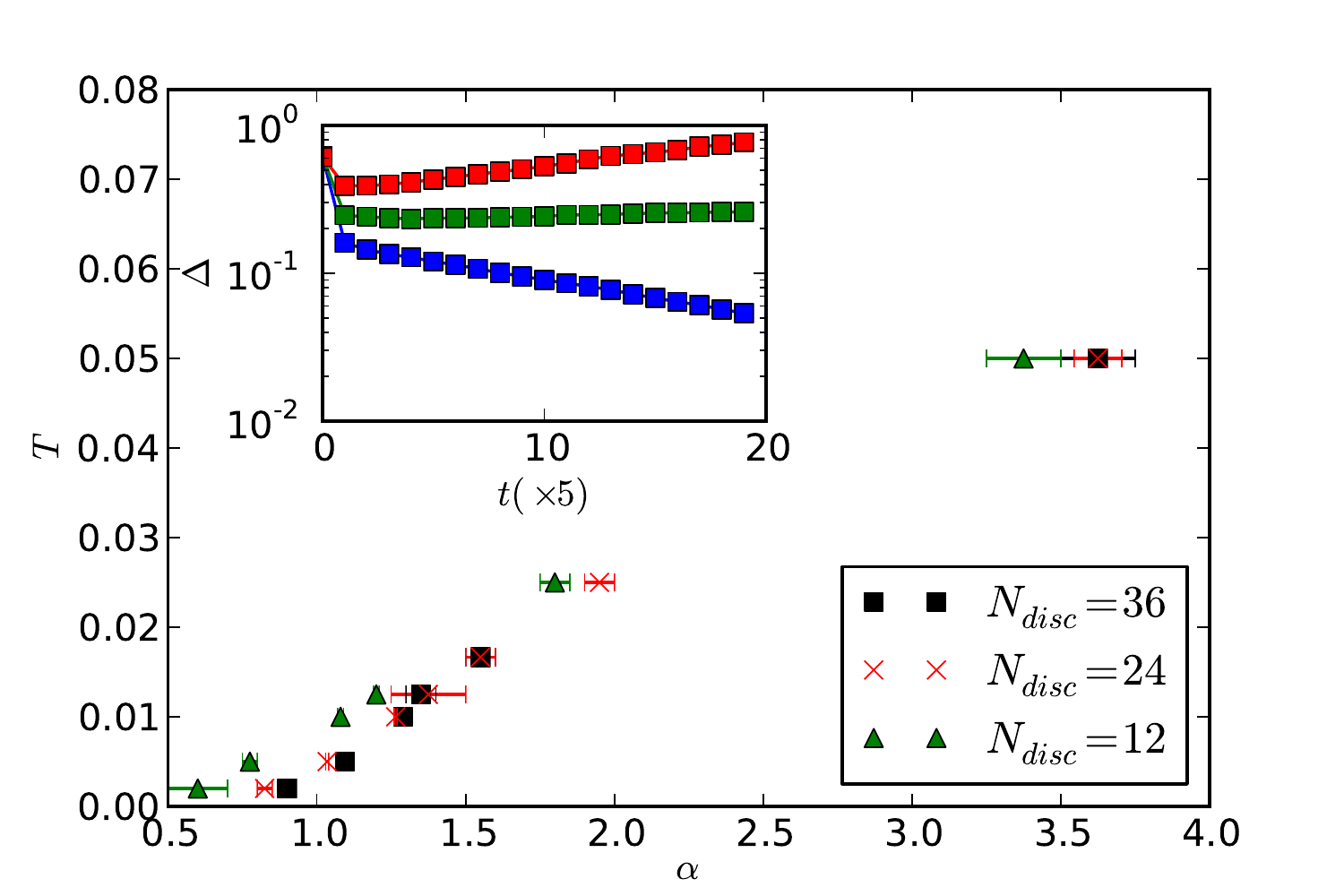}
	\caption{(inset) The sum-of-squares difference between two populations as a function of time in the population dynamics for $N_{disc}=36$ states and $\beta=100$. From bottom to top in the inset, we show data for $\alpha=1.26,1.28$ and $1.29$. We observed a transition in $\alpha$ between when the populations converge or diverge over time. This boundary is plotted in the main figure. 
	(main) The location of the dynamical instability phase boundary extracted from the Lyapunov exponents of the population dynamics. As the discretization of the Bloch sphere increases, the dynamical instability appears to converge to a finite temperature phase boundary. 
	As $T\rightarrow 0$, the instability line appears to terminate in the PRODSAT transition at $\alpha\simeq 0.91$. 
	\label{fig:lyap}}
\end{figure}

Fortunately, while the full treatment of replica symmetry breaking requires the solution of a yet more complicated hierarchy of equations than \eqref{eq:cavity-rs}, it is possible to detect the presence of replica symmetry breaking within the replica symmetric treatment \cite{Montanari-2003, Mertens-2003}.
To perform such a study, one initializes the population dynamics algorithm with two populations that differ by a small random displacement, $\delta \mathbb{P}$, and then evolves each population with an identical sequence of random moves. 
In the replica symmetric phase, these two initial conditions result in identical distributions for $P_{i\rightarrow m}$ and $Q_{m\rightarrow i}$ at late times. In the presence of replica symmetry breaking, the two distributions diverge. 
The Lyapunov exponent associated with this process (inset, Fig.~\ref{fig:lyap}) provides a robust indication of the underlying instability: where the exponent goes from negative to positive, we identify the dynamic instability phase boundary in the $(\alpha,T)$ plane (Fig.~\ref{fig:lyap}).  As can be seen best in Fig.~\ref{fig:energy-bdy}, the high energy, low clause density regime of  PSAT exhibits full replica symmetry while the dynamical instability sets in along a boundary which encloses the low energy density, high clause density regime, frustrating the approach to the ground state in the non-PRODSAT phase.

As our numerical approach relies on discretization of the continuous spins, we have attempted to check the convergence of the instability boundary to a finite limit as $N_{disc}$ becomes large. 
Indeed, for any discretization we find that there is an $\alpha_{c}(T)$ where there is a dynamical glass transition. 
As the discretization is increased, the instability line moves to higher clause density but, for the three discretizations probed, we believe this finite-discretization shift converges to a finite temperature boundary for $\alpha > 0.9$ (see Fig.~\ref{fig:lyap}). 
This is especially striking for, as $T\rightarrow 0$, we find that this transition apparently coincides with the zero temperature PRODSAT transition.\cite{Laumann:2010uq} 
We note that this is different from the situation in XORSAT where the dynamical glass transition at zero temperature coincides with the appearance of the hypercore at $\alpha_{hc} \simeq 0.81$ rather than the SAT-UNSAT transition at $\alpha = 0.91$ \cite{Altarelli:2009ta}.

\section{Concluding remarks} 
\label{sec:conclusion}

As noted at the outset this paper is fundamentally concerned with the approximation problem for randomly chosen QSAT instances. 
Specifically, by combining several
methods---cavity analysis, simulated annealing and a greedy algorithm---we give an estimate of the limits to which product states can be used to approximate solutions to 3-QSAT. Among these methods, the cavity analysis
shows a dynamical glass transition before the limits of product state
approximation are reached but this does not appear to have a marked 
influence on the performance of our best algorithm, the greedy search -- at least for sizes up to $N=100$. 
As such it is not clear whether the search for product states exhibits an
easy-hard transition before the limits of the product state region are
reached. Along the way, we have provided various additional statistical
mechanical results on the classical PSAT Hamiltonian which
may be of of interest to aficionados of disordered systems.

A natural next step is to study classical wavefunctions that build in 
some entanglement. It may be possible to analyze bounded-depth local quantum circuits matched to the random graph structure, similar to those studied in \cite{Hastings:2012tb}. It is unclear how the non-commuting nature of generic QSAT Hamiltonians affects the quality of this approach and whether one needs to move to more general entanglement structures.
We expect such considerations and the set of conjectures 
embodied in our opening figure (Fig.~1) to provide stimulus for future
work.


\begin{acknowledgments}
We thank Antonello Scardicchio and Francesco Zamponi for useful discussions.
This work was supported in part by NSF grant PHY-1005429 (BH, SLS).
CRL acknowledges support by a Lawrence Gollub Fellowship and the NSF through a grant for ITAMP at Harvard University.
BH and CRL thank the Max Planck Institut f\"ur Physik Komplexer Systeme for their hospitality while this work was being completed.
\end{acknowledgments}

\appendix

\section{Exact singletons} 
\label{sec:exact_singletons}

\subsection{Singleton clause} 
\label{sub:single_clause}

The partition function for a single $k=1$ clause attached to a spin is given by:
\begin{align}
	Z=\int d(\cos\theta)d\phi\, e^{-\beta |\phi_a z^a|^2}
\end{align}
Choosing the coordinate system to align with the state $\ket{\phi}$, this reduces to
\begin{align}
	Z&=\int d(\cos\theta)d\phi\, e^{-\beta \cos^2(\theta/2)} \\
	&=2\pi \int d(\cos\theta) e^{-\beta (1-\cos \theta)/2} \\
	&=2\pi e^{-\beta/2} \left. \frac{e^{\beta u/2}}{\beta/2} \right|_{-1}^1 \\
	&=4\pi  (1-e^{-\beta}) / \beta
\end{align}
The free energy is
\begin{align}
	F &= -\f{1}{\beta}\left[ \log 4\pi + \log\f{1-e^{-\beta}}{\beta}\right]
\end{align}
The entropy is
\begin{align}
	S &= - \f{\partial F}{\partial T} = \beta^2 \f{\partial F}{\partial \beta}\\
	&=\log 4\pi + \log\f{1-e^{-\beta}}{\beta} + 1 - \f{\beta}{e^{\beta} - 1}
\end{align}
The energy is
\begin{align}
	U &= F+S/\beta = \f{1}{\beta} - \f{1}{e^\beta-1}.
\end{align}

These expressions recover the correct limits (high temperature):
\begin{align}
	U&\sim \f{1}{\beta}(1-\f{1}{1+\beta/2+\beta^2/3!}) = \f{1}{2} - \beta/6 + O(\beta^2) \\
	S&\sim \log 4\pi + O(\beta^2)
\end{align}
and low temperature:
\begin{align}
	U &\sim 0 + \f{1}{\beta} + O(e^{-\beta}) \\
	S &\sim \log 4\pi/\beta + 1 + O(e^{-\beta})
\end{align}

To check the high temperature expansion, we have
\begin{align}
	\avg{H}_0 & = \avg{|\phi_a z^a|^2}_0= \avg{\f{1-u}{2}}_0 = \f{1}{2} \\
	\avg{H^2}_0 & = \avg{\f{1-2u + u^2}{4}}_0 = \f{1}{4}(1+ \f{1}{3}) = \f{1}{3} \\
	U &\sim \avg{H}_0 - \beta(\avg{H^2}^c_0) = \f{1}{2} - \beta\f{1}{12}
\end{align}
Here, we note that the infinite temperature measure on $u$ is simply uniform on $\left[-1,1\right]$.


\subsection{Single product clause} 
\label{sub:single_product_clause}

For a product projector, the integrals for the partition function may be simplified by rotating the local bases to match those of the projector. Thus,
\begin{align}
	Z &= \int d(\cos\theta_1)d\phi_1  \cdots d(\cos\theta_k)d\phi_k\, e^{-\beta |\phi^1_a z^a_1|^2 \cdots |\phi^k_b z^b_k|^2} \\
	&= (2\pi)^k\int du_1\cdots du_k e^{-\beta\f{1-u_1}{2}\cdots\f{1-u_k}{2}}
\end{align}
For $k=2$, this can be evaluated to the useless form,
\begin{align}
	Z = \f{(4\pi)^2}{\beta}\left[\gamma + \Gamma[0,\beta] + \log\beta\right]
\end{align}



\section{Uniform measure on $\mathbb{CP}^{n-1}$} 
\label{sec:uniform_measure}

The measure on the projectors is the uniform Haar measure on the space $\mathbb{CP}^{n-1}$. We can represent this measure on an $n$-component complex vector $\phi_i$
\begin{align*}
	Z &= \int
\left(\frac{d\phi_i^*d \phi_i}{2 i}\right)
\delta(|\phi|^2 - 1) \\
&= \int_{-\infty}^{\infty} \frac{d\lambda}{2\pi} \int
\left(\frac{d\phi_i^*d \phi_i}{2 i}\right)
e^{i \lambda (|\phi|^2 - 1)} \\
&= \int_{-\infty}^{\infty} \frac{d\lambda}{2\pi}
e^{-i \lambda}
\left(\frac{\pi}{-i \lambda}\right)^n\\
&= \left(\frac{\pi}{-i}\right)^n
\left.\int_{-\infty}^{\infty} \frac{d\lambda}{2\pi}
e^{-i \lambda k}
\frac{1}{\lambda^n}\right|_{k=1} \\
&= \left(\frac{\pi}{-i}\right)^n
\left.\int_{-\infty}^{\infty} \frac{d\lambda}{2\pi}
e^{-i \lambda k}
\frac{1}{(-1)^{n-1}(n-1)!} \partial^{n-1}_\lambda \frac{1}{\lambda}\right|_{k=1} \\
&= \left(\frac{\pi}{-i}\right)^n
\frac{1}{(-1)^{n-1}(n-1)!}
(ik)^{n-1}
\left.\textrm{sign}(k)\right|_{k=1} \\
&= \pi^n\frac{i}{(n-1)!}
\end{align*}

With the partition function in hand we can evaluate the disorder averaged correlators. The two-point correlator is
\begin{align}
	\overline{\phi^*_i \phi_j} &=
	\frac{1}{Z}
	\int_{-\infty}^{\infty} \frac{d\lambda}{2\pi} \int
	\left(\frac{d\phi_i^*d \phi_i}{2 i}\right) \phi^*_i \phi_j
	e^{i \lambda (|\phi|^2 - 1)} \nonumber\\
	&=
	\frac{1}{Z}
	\int_{-\infty}^{\infty} \frac{d\lambda}{2\pi}
	e^{-i \lambda}
	\left(\frac{\pi}{-i \lambda}\right)^n \frac{\delta_{ij}}{-i \lambda}\nonumber\\
	& = \frac{\delta_{ij}}{n}
\end{align}
The higher order correlators satisfy a slightly modified Wick's theorem:
\begin{align}
	\overline{\phi^*_i \phi^*_j \phi_k \phi_l} & = \frac{\delta_{ik}\delta_{jl} + \delta_{il}\delta_{jk}}{n(n+1)} \\
	\overline{\phi^*_1 \phi^*_2 \cdots \phi^*_k \phi_{k+1} \phi_{k+2} \cdots \phi_{2k}} & = \frac{\sum_{pairings} \delta \delta \cdots \delta}{n(n+1)\cdots(n+k-1)}
\end{align}

These averages are equally useful for evaluating high temperature correlators of the two-component spinors $z_a$.


\section{Hypercore data} 
\label{sec:hypercore_datum}

We summarize a few results derived in \cite{Mezard:2003p5977} regarding the hypercore of random factor graphs $G$. The hypercore is the maximal subgraph of $G$ on which all nodes have degree at least 2. The quoted results follow from analyzing the leaf removal algorithm applied to the random graph. In the notation of \cite{Mezard:2003p5977}, the clause density $\alpha$ is $\gamma$ and the number of spins per clause $k$ is $p$.

The degree distribution on the hypercore is Poissonian for degrees $d\ge 2$:
\begin{align}
	P_c(d) = \left\{ \begin{array}{ll} 0 & \textrm{for } d=0,1\\
	e^{-\lambda^*(\alpha)}\f{\lambda^*(\alpha)^d}{d!} & \textrm{for } d\ge 2 \end{array} \right.
\end{align}
where the average degree $\lambda^*$ corresponds to the largest solution of the equation:
\begin{align}
	e^{-\lambda^*}-1+\left(\f{\lambda^*}{k\alpha}\right)^{\f{1}{k-1}} = 0.
\end{align}

The total number of nodes in the hypercore is
\begin{align}
	N_c(\alpha) = N \sum_{d \ge 2} P_c(d) = N\left[ 1-(1+\lambda^*)e^{-\lambda^*}\right]
\end{align}
and the total number of edges (clauses)
\begin{align}
	M_c(\alpha) = N\f{\lambda^*}{k}(1-e^{-\lambda^*}).
\end{align}
For large $\alpha$, the hypercore takes over most of the graph up to exponentially small corrections: $\lambda^* \approx k \alpha$ and $N_c = N(1- (1+k \alpha)e^{- k \alpha})$, $M_c = N \alpha(1-e^{-k\alpha})$.


%

\end{document}